\documentstyle[preprint,aps]{revtex}
\tightenlines
\def\beq{\begin{equation}}
\def\eeq{\end{equation}}
\begin{document}
\title{Comparison of Relativistic Nucleon-Nucleon Interactions}
\author{T. W. Allen, G. L.  Payne, Wayne N. Polyzou}
\address{ 
Department of Physics and Astronomy, 
The University of Iowa, 
Iowa City, IA 52242
}
\date{\today}
\maketitle
\footnotetext{P.A.C.S. 03.65.Pm, 21.30.x,21.45.+v, 21.10.Jv,25.10.+s,  
25.30.Bf} 
\begin{abstract}

We investigate the difference between those relativistic 
models based on interpreting
a realistic nucleon-nucleon interaction as a perturbation of the square of a
relativistic mass operator and 
those models that use the method of Kamada and
Gl\"ockle to construct an equivalent interaction to add to the
relativistic mass operator.  Although both models reproduce the phase 
shifts and binding energy of the corresponding non-relativistic model, 
they are not scattering equivalent.  The example of elastic electron-deuteron 
scattering in the one-photon-exchange approximation is used to study the 
sensitivity of three-body observables to these choices.  Our conclusion is 
that the differences in the predictions of the two models can be understood 
in terms of the different ways in which the relativistic and non-relativistic
$S$-matrices are related. We 
argue that the mass squared method is consistent with conventional 
procedures used to fit the Lorentz-invariant cross section as a function of 
the laboratory energy.

\end{abstract}
\pacs{03.65.Pm, 21.30.x,21.45.+v, 21.10.Jv,25.10.+s,  25.30.Bf}
\newpage
\narrowtext
\section{Introduction}

Realistic relativistic quantum mechanical models of interacting nucleons 
are important for modeling medium and high-energy reactions 
involving nuclear targets.  The simplest 
generalization of the non-relativistic two-body Hamiltonian is 
\beq
h={k^2 \over m} + V \qquad \to \qquad M= 2\sqrt{k^2 + m^2} + V'.
\label{eq:rocco}
\eeq 
In this case the mass operator, $M$, replaces the non-relativistic center of 
momentum Hamiltonian, $h$.  Bakamjian and Thomas \cite{bakamjian} have 
determined sufficient conditions on $V'$ in (\ref{eq:rocco}) so $M$ 
can be interpreted as the mass operator of an equivalent relativistic 
quantum model.

In the non-relativistic case, interactions are determined by fitting parameters
in the interaction so that the solutions 
of the dynamical equations reproduce the
existing bound-state and scattering observables.  This is a non-trivial task that 
has led to the construction of a number of realistic
interactions \cite{ried}\cite{wiringa}
\cite{bonn}.  Schiavilla \cite{rocco} has repeated this analysis,
by refitting parameters in an Argonne V14 interaction,  to 
determine a realistic relativistic interaction, $V'$.  It is tedious to 
have to refit parameters for every new interaction.

A direct way to utilize existing non-relativistic interactions, 
without refitting the interaction,  is preferable.  Two methods exist that
can be used to directly formulate relativistic models based on realistic 
non-relativistic interactions.  Historically the first of these methods 
was introduced by Coester, Pieper, and Serduke \cite{serduke}
\cite{lee}\cite{coester}.  In their 
construction an interaction is added to the square of the non-interacting 
mass operator.  The interaction is chosen so the dynamical equations are
equivalent to the non-relativistic equations.  The result is that the 
eigenfunctions of the relativistic model can be trivially obtained
from those of the corresponding non-relativistic model. 
The second method was introduced by Kamada and Gl\"ockle \cite{kamada}.  
This utilizes an $S$-matrix preserving unitary transformation, called a 
scattering equivalence \cite{coester0}\cite{saenz}\cite{coester}, that 
relates the 
two operators in (\ref{eq:rocco}).  The transformation involves a simple 
rescaling and a renormalization that ensures unitarity.  Both methods 
lead to invariant $S$-matrices that are identical to the non-relativistic 
$S$-matrix.

In this paper we investigate the difference between the mass-squared method and
the Kamada-Gl\"ockle method by comparing both the two-body dynamics and
elastic electron-deuteron scattering, which is the simplest three-body 
reaction.

The surprising conclusion of our analysis is that even though both methods
give the same invariant scattering matrix as the corresponding 
non-relativistic model,  they do not lead to identical predictions, 
even at the two-body level.  Specifically, these two methods do 
not lead to scattering equivalent theories.   

We also consider differences in the predictions for elastic 
electron-deuteron form factors based on these two models.  While normally 
a change in the two-body interaction would lead to a corresponding
change in the strong current,  in typical calculations the consistency of
the dynamics and the strong current is normally limited to ensuring 
current conservation and current covariance.

The comparisons in this paper are limited to understanding the 
quantitative sensitivity of the physical observables to the 
two different models of the two-body dynamics, keeping everything 
else the same.
Specifically we 
compare the calculated observables using the deuteron eigenstates 
constructed from the mass-squared and Kamada-Gl\"ockle methods based 
on the same realistic nucleon-nucleon interaction.  In both calculations 
the current is treated in an identical impulse-like approximation, using 
identical nucleon form factors and an identical form of the dynamics.
It is impossible to rigorously determine if the difference in 
these three-body observables are differences that would arise if the two 
methods were two-body scattering equivalent, or if they are due to 
the violations of the scattering equivalence.  However,  the result of
the above comparison is that there are small but measurable differences
in the two predictions.  The parameter that controls the scale of these 
differences is the same one that controls the scale of the violation of the 
scattering equivalence.

For the electron-scattering calculations the coupling to the electron 
is treated in the one-photon-exchange approximation.  The relativistic 
models are formulated in Dirac's point form of the dynamics.  The hadronic 
current operators are modeled using the point-from spectator approximation 
\cite{allen}.  The input nucleon current matrix elements are constructed 
from the Gari-Kr\"umpelmann parameterization\cite{gari}.  The specific 
choice of realistic interaction, form of dynamics, or nucleon form factors 
is not central to our analysis.

The notion of scattering equivalence is necessary to understand 
the issues involved in our analysis.  Unitary transformations 
preserve probabilities in quantum mechanics.
The conventional interpretation of this observation is that the
physical predictions of a quantum theory do not depend on the choice
of Hilbert space basis.

There is a more subtle interpretation of this result that is best
understood in the context of scattering theory.  The S-matrix measures
the difference between the free and interacting Hamiltonian that can be 
observed asymptotically.   Scattering observables are preserved under 
unitary transformations provided the same unitary transformation is 
used to transform both the free and interacting Hamiltonians.

A class of unitary transformations, called scattering equivalences,
have the property that if they are applied to the full Hamiltonian
{\it without} applying them to the free Hamiltonian, they give the same
scattering matrix elements and bound state observables as the original
Hamiltonian.  They are not meant to transform representations of free 
particles.  In this sense they have a different interpretation than a 
standard unitary transformation.  They parameterize the insensitivity of 
the asymptotic $S$-matrix observables to certain short-range features
of the interactions.

Scattering equivalences are unitary transformations $W$
that satisfy the asymptotic condition\cite{saenz}\cite{coester}\cite{glockle1}
\beq
\lim_{t \to \pm \infty} \Vert (I-W)e^{-iH_0t}\vert \Phi \rangle \Vert =0.
\eeq
Examples of scattering equivalences include operators of the form
\beq
W= {I+ i C \over I-iC} 
\eeq
where $C$ is any compact Hermitian operator.

In a field like nuclear physics, where model interactions are determined by
fitting to $S$-matrix and bound-state observables, realistic model Hamiltonians
can at best be determined up to an overall scattering equivalence.  The
ambiguity in the choice of representation of the interaction is particularly
confusing in the case of relativistic models, where there are many scattering
equivalent ways to interpret a given non-relativistic interaction.  
For example,  the kernel of a single two-body interaction defines 
different operators in each of Dirac's \cite{dirac} forms of the dynamics.  
The operators differ by whether they commute with the total linear momentum, 
the total four velocity, or the light-front components of the four momentum. 
The resulting two-body models are scattering equivalent, but lead to 
different three-body predictions.  Even within a single form of dynamics, 
there are scattering equivalences \cite{polyzou} that relate different 
natural choices of two-body interactions.  In all of these cases, the 
different scattering equivalent models have the same two-body bound state and
scattering observables.  They can be made to have the same three-body
observables by introducing additional three-body interactions\cite{glockle1}.
If, as is customary, three-body interactions are not included, then different
scattering equivalent two-body models give different three-body predictions.

In what follows a brief description of each model is given.  Differences 
in the $S$-matrix,  wave functions, and elastic electron-deuteron 
observables are analyzed.

\section{The mass-squared method:}

In this section the mass-squared method is described for the  
case of two particles with equal masses.  The case of unequal 
masses is treated in \cite{keister}. 

Let $V_{nr}$ be a non-relativistic nucleon-nucleon interaction
defined by a kernel of the form
\[
\langle {\bf p} , k , j, \sigma , l , s \vert V_{nr} \vert
{\bf p}\,' , k' , j' , \sigma' ,l' ,s' \rangle 
\]
\beq
=\delta ({\bf p}- {\bf p}\,') \delta_{jj'} \delta_{\sigma \sigma'} 
\langle k , l , s \Vert v_{nn}^j \Vert k' , l' , s' \rangle,  
\label{eq:BA}
\eeq
where $k$ is the magnitude of the relative momentum.
The physics is contained in the reduced kernel
\beq
\langle k , l , s \Vert v_{nn}^j \Vert k' , l' , s' \rangle . 
\label{eq:BB}
\eeq	

Define a model relativistic mass operator:
\beq
M^2 := 4(k^2 + m^2) + 4mV_{m^2};
\label{eq:BC}
\eeq
where $V_{m^2}$ is a point-form interaction defined by
\[
\langle {\bf v} , k , j, \sigma , l , s \vert V_{m^2} \vert
{\bf v}\,' , k' , j' , \sigma' ,l' ,s' \rangle 
\]
\beq
= \delta ({\bf v}- {\bf v}\,') \delta_{jj'} \delta_{\sigma \sigma'} 
\langle k , l , s \Vert v_{nn}^j \Vert k' , l' , s' \rangle.  
\label{eq:BD}
\eeq
The interaction kernel of $V_{m^2}$  is identical to the 
non-relativistic kernel (\ref{eq:BB}).  In this expression ${\bf v}$ 
is the velocity of the 
non-interacting two-body system, $k$ is related to the invariant mass 
of the non-interacting system by $M_0^2 = 4(k^2 + m^2)$, and $j$ is 
the spin of the non-interacting two body-system.  

The two-body eigenvalue problem, 
\beq
M^2 \vert \psi \rangle = \lambda^2 \vert \psi \rangle , 
\label{eq:BE}
\eeq
is mathematically equivalent to the non-relativistic Schr\"odinger equation
\beq
H_{nr} \vert \psi \rangle = 
({k^2 \over m} + V_{nr}) \vert \psi \rangle = -\epsilon_{nr} 
\vert \psi \rangle .
\label{eq:BF}
\eeq
The eigenfunctions are identical functions.  Because of the relation
\beq
M^2= 4m(H_{nr} +m),
\eeq
the eigenvalues of these equations are related by
\beq
\lambda^2 = (2m-\epsilon_r)^2 =  4m(-\epsilon_{nr} +m ).
\eeq
It follows that for the deuteron
\beq
{\epsilon_{r} - \epsilon_{nr} \over \epsilon_r}=
{\epsilon_r \over 4m} \sim {1 \over 2000},
\eeq
which implies that the relativistic and non-relativistic binding energies
are identical to within experimental uncertainties.  

The two-body mass operator (\ref{eq:BC}) also leads to the same cross 
section as in the non-relativistic model (\ref{eq:BF}). 
To see this note that the Lorentz invariant \cite{moller} relativistic 
cross section is given by
\[
\left ({d\sigma \over d\Omega d\epsilon} \right )_{r} 
\]
\[
={(2 \pi)^4 \over v_r }
\vert \langle f^+ \vert M-M_{0} \vert i 
\rangle \vert^2 k^2 {dk \over d\epsilon_r}
\]
\[
={(2 \pi)^4 \over 2k/\omega }
\vert \langle f^+ \vert M-M_{0} \vert i \rangle \vert^2 k^2 
{\omega \over 2k}    
\]
\[
=(2 \pi)^4
\vert \langle f^+ \vert {M^2-M^2_{0}\over 4 \omega} \vert i \rangle 
\vert^2  \left ( {\omega \over 2} \right )^2 
\]
\[
=(2 \pi)^4
\vert \langle f^+ \vert V_{m^2} {m \over \omega} \vert i \rangle 
\vert^2  \left ({\omega \over 2} \right )^2 
\]
\[
={(2 \pi)^4 \over 2k/m }
\vert \langle f^+ \vert V_{m^2} \vert i \rangle \vert^2 k^2 
{m \over 2k}
\]
\beq
{(2 \pi)^4 \over v_{nr} }
\vert \langle f^+ \vert V_{nr} \vert i 
\rangle \vert^2 k^2 {dk \over d\epsilon_{nr}};   
\label{eq:CX}
\eeq
where the relativistic and non-relativistic velocities are
\beq
v_r={2k \over \omega},  \qquad \omega = \sqrt{k^2+m^2}, \qquad v_{nr} = 
{2k \over m};
\eeq
and the relativistic and non-relativistic kinematic internal energies 
are 
\beq
\epsilon_r = 2 \omega, \qquad \mbox{and} \qquad \epsilon_{nr} = 2m + {k^2 \over m}. 
\eeq
The calculation in (\ref{eq:CX}) shows that the relativistic cross section 
based on $M^2$ is identical to the non-relativistic cross section 
derived from the non-relativistic Hamiltonian (\ref{eq:BF}).

This shows that the relativistic invariant cross section (with the 
relativistic velocity and phase space factor) is identical to the 
non-relativistic cross section (with the non-relativistic velocity 
and phase space factor).  The two-body eigenstates of the non-relativistic 
Hamiltonian (\ref{eq:BF}) are also eigenstates of the relativistic 
mass operator (\ref{eq:BC}).  

In this case, because the factors $k^2$ in equations (\ref{eq:BC}) and
(\ref{eq:BF}) are identified, the scattering matrix elements are
identified as functions of invariant relative momenta $k$:
\beq
S_r(k)=S_{nr}(k).
\label{eq:SA}
\eeq
These momenta have different interpretations in the relativistic and 
non-relativistic cases when they are expressed in terms of single 
nucleon degrees of freedom.  Both correspond to the same asymptotic 
observable,  the momentum of a single particle in the center of 
momentum frame of the non-interacting system.

Ignoring the  difference in relativistic and non-relativistic binding 
energy,  the mass squared model is scattering equivalent to the 
non-relativistic model.  There is no need to refit interaction parameters.

\section{The Kamada-Gl\"ockle interaction}

\bigskip
The Kamada-Gl\"ockle method is an alternate method that is used to make 
relativistic quantum models using a realistic nucleon-nucleon model as input.

The Kamada-Gl\"ockle method defines the operator $q$ by 
\beq 
2\sqrt{m^2 + k^2} := 2m + {q^2 \over m} = F(q).
\label{eq:kin}
\eeq
Equation (\ref{eq:kin}) can be solved for  $q(k)$ and $k(q)$:
\beq
q^2 = m ( 2\sqrt{m^2 +k^2} -2m);
\label{eq:kq}
\eeq
\beq
k^2 = q^2 \left ( 1 + {q^2 \over 4m^2} \right ).
\eeq
Kamada and Gl\"ockle also define $h(q)$ by the equation
\beq
k^2 dk = h^2 (q) q^2 dq
\label{eq:CA}
\eeq
which can be solved to give
\beq
h^2(q) = {k \over q}\left ( 1 + {q^2 \over 2m^2}\right ). 
\eeq
Defining the operator $V_{kg}$ by
\[
\langle {\bf v} , k , j, \sigma , l , s \vert V_{kg} \vert
{\bf v}\,' , k' , j' , \sigma' ,l' ,s' \rangle 
\]
\beq
= \delta ({\bf v}- {\bf v}\,') \delta_{jj'} \delta_{\sigma \sigma'} 
{1 \over h(q(k))}\langle k , l , s \Vert v_{nn}^j \Vert k' , l' , s' \rangle  
{1 \over h(q(k'))}
\label{eq:CD}
\eeq
where $\langle k , l , s \Vert v_{nn}^j \Vert k' , l' , s' \rangle$ is the 
reduced interaction (\ref{eq:BA}),
Kamada and Gl\"ockle show that the operator  
\beq
M= 2 \sqrt{k^2 + m^2} + V_{kg} - 2m 
\eeq
has the exact same deuteron binding energy as predicted by the
non-relativistic model.  In addition, they show that the $S$-matrix
elements predicted in their model and the
non-relativistic model, as a function of
\beq
\epsilon = 2\sqrt{m^2 + k^2}-2m  = {q^2 \over m},
\eeq
are identical.  Thus, these two models are scattering equivalent.

The relativistic wave functions and transition matrix elements can
be obtained directly from the corresponding non-relativistic quantities: 
\beq
\phi^j (k,l,s) = \psi^j (q(k),l,s)/h(q(k)) 
\eeq
and 
\beq
T^j_r (k,l,s; \epsilon; k',l',s') := {1 \over h(q(k))} T^j_{nr} (q(k),l,s ; \epsilon ;
q(k'),l',s') {1 \over h(q(k'))}.
\eeq
This has the advantage that the solution of the relativistic scattering and 
eigenvalue problems do not have to be recalculated.

With this method, the non-relativistic and relativistic scattering
matrix elements are 
identified as identical functions of the non-relativistic and relativistic 
invariant center of momentum energy: 
\beq
S_r(\epsilon)=S_{nr}(\epsilon);
\eeq
where
\beq
\epsilon= {q^2/m} + 2m = 2\sqrt{k^2 +m^2}.
\eeq
Thus,  the Kamada-Gl\"ockle model is also scattering
equivalent to the non-relativistic model. 
 
\section{Discussion}

Each of the methods discussed above provides a simple prescription for
interpreting existing realistic non-relativistic models as equivalent
relativistic models.  In the mass-squared  case a local interaction is 
added to the square of the free mass operator and in the 
Kamada-Gl\"ockle case a non-local operator is added to the mass operator. 

Because each model has a relativistic scattering matrix that is
equivalent to the non-relativistic scattering matrix, one might expect
that the two relativistic models would be scattering equivalent.  This
is {\it not} true.  This is because in the mass-squared case the
$S$-matrix elements are identified as functions of $k$ while in the
Kamada-Gl\"ockle case they are identified as function of the invariant
energy.

In order to get a quantitative understanding of the difference in the 
$S$-matrix elements of these two models, note that both models solve 
the Schr\"odinger equation:
\beq
({q^2 \over m} + V_{nr}) \vert \phi \rangle = 
\epsilon_{12} \vert \phi \rangle
\qquad 
({k^2 \over  m} + V_{m^2}) \vert \phi \rangle = \epsilon_{12} 
\vert \phi \rangle.
\label{eq:eq}
\eeq
Although the operators are different, the internal wave functions are the 
non-relativistic wave function as a function of $q$ or $k$ respectively.
It follows that the corresponding scattering matrix elements are related by
\beq
S_{KG}(\epsilon(k_{KG})) = S_{nr}(\epsilon_{nr}(q)) = 
S_{m^2}(k_{m^2}) \qquad \mbox{for } \qquad k_{m^2}=q.
\label{eq:DA}
\eeq
The problem is that at point where all of the $S$-matrix elements agree,
we have $k_{m^2} = q \not= k_{KG}$.  
On the other hand, $k_{KG}$ and $k_{m^2}$ represent the same asymptotic
observable in the two relativistic models, which means the the physics 
requires 
\beq
S_{m^2} (k_{m^2}) = S_{KG} (k_{KG}) \qquad \mbox{for} \qquad
k_{m^2} = k_{KG}
\eeq
which contradicts (\ref{eq:DA}).

To show this explicitly note by definition
\beq
k_{KG}^2 = q^2 (1 + q^2/4m^2).
\label{eq:kinc}
\eeq
The point where the mass squared and Kamada-Gl\"ockle $S$-matrices agree 
corresponds to $k_{m^2} = q$.  Substituting $k^2_{m^2}$ for $q^2$ in the 
above equation gives the following relation between $k_{m^2}$ and 
$k_{KG}$:
\beq
k_{KG}^2 = k_{m^2}^2 (1 + k_{m^2}^2/4m^2). 
\label{eq:kinb}
\eeq
Since
$k_{KG} > k_{m^2}$ at the point where the $S$-matrix elements agree, this
implies that the $KG$ cross sections fall off slower than the mass-squared 
cross
sections when considered as functions of $k$.

There is nothing theoretically wrong with either of these methods; 
nevertheless they are inconsistent with each other and cannot both 
be consistent with experiment. 
 
The key to resolving this problem was originally noted by Breit 
\cite{breit}.  He correctly pointed out that the laboratory cross 
section is a Lorentz invariant function of a Lorentz invariant 
variable, the laboratory energy.  Specifically, the laboratory 
energy $\epsilon_{lab}$ can be expressed in a manifestly invariant 
manner by:
\beq
p_1 \cdot p_2 = -\epsilon_{lab} m = - m^2 - 2k^2 ;
\eeq
or
\beq
\epsilon_{lab} =  2{k^2\over m} + m, \qquad 
k^2 = {m \over 2} (\epsilon_{lab} -m) .
\eeq
This indicates that the experimentally determined $S$-matrix
and cross sections are Lorentz invariant functions of the 
invariant $\epsilon_{lab}$ or equivalently $k^2$. 
The key point is that there is no such thing as the experimentally determined
non-relativistic cross section.  

``Non-relativistic'' potentials are determined by adjusting 
the parameters of the potential until the scattering eigenstates
of 
\beq
H= {k^2 \over m} +V
\eeq 
reproduce the experimentally measured $S(k^2)=S(\epsilon_{lab})$.  The
analysis of the mass-squared method in equation (\ref{eq:CX}) shows that the
non-relativistic and relativistic formulas for the cross section lead
to the same invariant quantity when $k^2$ is identified with the
asymptotic relative momentum in the center of momentum frame.

If this same data were used to determine the Kamada-Gl\"ockle interaction, 
it would be appropriate to determine the interaction $V_{KG}$ directly 
by comparing to the Lorentz invariant cross section as a function of $k^2$,
as was done by Schiavilla \cite{rocco}.  
The identification (\ref{eq:kin}) defines $q$ by
\beq
q^2=m(2 \sqrt{k^2_{m^2} + m^2} -2m)
\eeq
rather than by $k_{m^2}$.  This would lead to a ``non-relativistic'' 
model with a different high $q$ behavior.

In the case of the Kamada-Gl\"ockle method, because the measured 
invariant cross section is given as a function of $k^2_{KG}$ rather 
than $q^2$, the fitting should be done as a function of $k^2$, 
as in the mass-squared method, with the corresponding $q$ dependence
determined by the identification 
\beq
q^2 = m(2\sqrt{m^2 + k_{KG}^2} - 2m)=
m(2\sqrt{m^2 + {m\over 2}(\epsilon_{lab}-m)} - 2m).
\eeq

This suggests that a Kamada-Gl\"ockle interaction constructed from a
non-relativistic model fit to the $S$-matrix as a function of laboratory energy
will not be consistent with experiment at high energies.  In practice, the
differences are negligible as long as $k^2/4m^2$ is small.  For momenta where
this quantity is large there are questions about the interpretation of
non-relativistic models.

Although the wave functions in the two models are not observable, they are
shifted in the same manner as the $S$-matrix. The rescaling and 
renormalization of the wave function is
responsible for all of the differences in 
our computations of the electron scattering observables.  The result 
is that the Kamada-Gl\"ockle wave functions will be spread out
more in momentum space than the mass-squared deuteron wave functions.

The factor $k^2/4m^2$ in (\ref{eq:kinb}) defines the scale of the
shift of the $S$-matrix or wave function at a given value of $k$ that
results from the two different interpretations of the $k$ variables.  This
factor also defines the scale at which the two models fail to be
scattering equivalent and the scale on how much the the factor
$h(q^2)$ in (\ref{eq:CA}) differs from unity.  This suggests that the
differences in the wave functions of the two models are not simply due to 
an underlying scattering equivalence.  Instead, they are
completely determined by the difference in the momentum dependence of the 
$S$-matrices.  In the case of the electron-scattering predictions,
using these wave functions in the same point-form spectator 
approximation leads to 
differences in the elastic observables.  While this in not a true test of 
the three-body sensitivity, this comparison is consistent with the 
way that models of the hadronic current are tested.

Figures 1 and 2 show the momentum space $s$ and $d$ components of the 
deuteron wave functions for 
the mass-squared and Kamada-Gl\"ockle models based on the same Argonne
V18 potential.  These are plotted as functions of the variable $k$, which 
is a function of invariant mass of two free particles.  In the mass-squared 
case this is also interpreted as the non-relativistic relative momentum,
while in the Kamada-Gl\"ockle method the corresponding ``non-relativistic'' 
relative momentum, $q$, is smaller.  As discussed above, this tends to spread 
out the momentum space Kamada-Gl\"ockle wave functions relative to the 
mass-squared wave functions when plotted as functions of $k$.   This is 
seen in the plots of the $s$ and $d$-state wave deuteron wave functions in 
Figures 1 and 2.

Figure 1 shows the reduced $s$-state wave function and Figure 2 shows 
the reduced $d$-state wave function.  The solid line in both figures 
corresponds to the mass-squared model while the dashed line corresponds to 
the Kamada-Gl\"ockle model.  Both figures show the predicted shift of 
the Kamada-Gl\"ockle wave function to the right  of the mass-squared wave
function.

The shift is due to the relationship between $k$ and $q$ in (\ref{eq:kinc})
rather than due to the factor $h(q)$, which is a compensating factor 
that simply ensures that the change of variables is unitary.  Because 
this shift is minimal when $q^2/4 m^2 \ll 1$, the two wave functions 
look similar for momenta out to about $5$fm$^{-1}$ ($q^2/4m^2 \sim .25$).

We conclude that the difference in these two models is due to the 
difference in how the non-relativistic invariant energy is interpreted
relativistically.  Specifically with the mass-squared method, $k^2$ is 
identified as a relativistic relative momentum while in the Kamada-Gl\"ockle
method the relativistic and non-relativistic rest energies are identified.
These differences explain how the models fail to be 
scattering equivalent.  Nevertheless, both models 
provide reasonable relativistic interpretations of non-relativistic models.
They differ on scales where the non-relativistic interpretation is in 
question. 

These differences have implications for the three-body problem.
Because current matrix elements can couple 
high and low momentum components of the wave functions, the  
possibility of observable differences between these two approaches 
at momentum transfers below 1(GeV)$^2$ exists.  This is particularly relevant 
in the point-form spectator approximation, where the momentum transfer seen 
by a constituent nucleon is larger than the momentum seen by the 
deuteron \cite{allen} 

The calculations shown in Figures 3, 4, and 5 are the deuteron elastic electron
scattering observables $A(Q^2)$, $B(Q^2)$, and $T_{20}(Q^2)$ respectively 
in the 
point-form spectator approximation \cite{allen}.  In the point-form
spectator approximation matrix elements of the transverse 
current,
\beq
\tilde{J}^{\mu} (0) := J^{\mu} (0) - q^{\mu} {q \cdot J (0) \over q^2} ,
\eeq
are identified with matrix elements of the one-body operator,
\beq
\tilde{J}^{\mu} (0)= \tilde{J}_n^{\mu} (0)+  \tilde{J}_p^{\mu} (0),
\eeq
in the Breit frame.  All of the other current matrix 
elements are determined by 
current conservation and
current covariance.  The one-body currents are determined using
Gari-Kr\"umpelmann \cite{gari} single-nucleon form factors. 
The dotted curves
are the the result of using the Argonne V18 interaction to construct a
Kamada-Gl\"ockle interaction and solid curves are the result of using the same
interaction to construct a mass-squared interaction.  

The results show small, but measurable differences, in $A(Q^2)$ at 
momentum transfers near 1 (Gev)$^2$.  The observables $B(Q^2)$ exhibit 
differences in the two 
predictions that extend below the scales where differences are expected.
The data in Figure 3 is
from refs. \cite{elias}\cite{arnold1}\cite{martin}\cite{simon} while the 
data in Figure 3 is from \cite{buchanan}
\cite{galster}\cite{simon}\cite{arnold2}.

The results for the tensor polarization, $T_{20}(Q^2)$,
are also similar.  Differences in the predictions of 
the two methods again extend below $q^2= 1$(GeV)$^2$.  
The data are from \cite{schulze}
\cite{gilman}\cite{the}\cite{garcon}\cite{ferro}.

We conclude that the difference in these two methods is due to differences in
the interpretation of the relative momentum variable that appears in the
non-relativistic Hamiltonian.  The relevant rescaling is given in
(\ref{eq:kinb}).  The variable $k^2/4m^2$ defines a scale where the different
interpretations of the momenta become significant.  This is the scale on which
the two methods fail to be scattering equivalent and is the scale that defines
the shift in the momentum space wave functions.  The mass-squared
method is consistent with parameterizations of the Lorentz invariant 
cross section as a function of
the projectile energy in the laboratory frame.  For a fixed current model, the
three-body calculations are sensitive to the differences in the wave functions.
The model calculations show that these differences can have a measurable impact
at much lower momentum transfers.  To consistently interpret the 
Kamada-Gl\"ockle model relativistically, the non-relativistic phase 
shifts would have
to be refit.  This only impacts the high momentum structure of the
model.

These differences are shown to have measurable implications in elastic 
electron-deuteron scattering.  In the realistic model presented, these differences lead
to measurable effects in the observables $B(Q^2)$ and $T_{20}(Q^2)$ 
for momentum 
transfers as low as a few tenths of a (GeV)$^2$, far lower than the scales 
where differences are expected.  

We conclude that the mass-squared and Kamada-Gl\"ockle methods define 
different dynamical models and that the mass-squared method is the one that 
is consistent with conventional methods for determining experimental 
cross sections.   Quantitively, the differences in these models are small, 
but they have dynamical consequences at surprisingly low momentum 
transfers in the three-body system.

This work was supported by the Department of Energy, Nuclear
Physics Division, under contract DE-FG02-86ER40286.

\pagebreak
\mediumtext

\vfill\eject
\centerline{FIGURES}
\bigskip
\begin{enumerate} 
\item[1.] FIG. 1. The momentum space deuteron s-wave wave function 
for a relativistic point-form dynamics based on the 
mass squared model (solid line) and the Kamada-Gl\"ockle model 
(dashed line).
\bigskip
\item[2.] FIG. 2. The momentum space deuteron d-wave wave function 
for a relativistic point-form dynamics based on the 
mass squared model (solid line) and the Kamada-Gl\"ockle model (dashed line).
\bigskip
\item[3.] FIG. 3. $A(Q^2)$ for the mass squared interaction (solid line)
and the Kamada-Gl\"ockle model (dashed line).  Both calculation use Argonne 
V18 as the input interaction, Gari-Kr\"umpelmann \cite{gari} form factors,
and the point-form spectator approximation.  Data is from \cite{elias}
\cite{arnold1}\cite{martin}\cite{simon}. 
\bigskip
\item[4.] FIG. 4. $B(Q^2)$ for the mass squared interaction (solid line)
and the Kamada-Gl\"ockle model (dashed line).  Both calculation use Argonne 
V18 as the input interaction, Gari-Kr\"umpelmann \cite{gari} form factors,
and the point-form spectator approximation.  Data is from \cite{buchanan}
\cite{galster}\cite{simon}\cite{arnold2}.
\bigskip
\item[5.] FIG. 5 $T_{20}(Q^2)$ for the mass squared interaction (solid line)
and the Kamada-Gl\"ockle model (dashed line).  Both calculation use Argonne 
V18 as the input interaction, Gari-Kr\"umpelmann \cite{gari} form factors,
and the point-form spectator approximation.  Data is from \cite{schulze}
\cite{gilman}\cite{the}\cite{garcon}\cite{ferro}. 
\bigskip
\end{enumerate}
\vfill
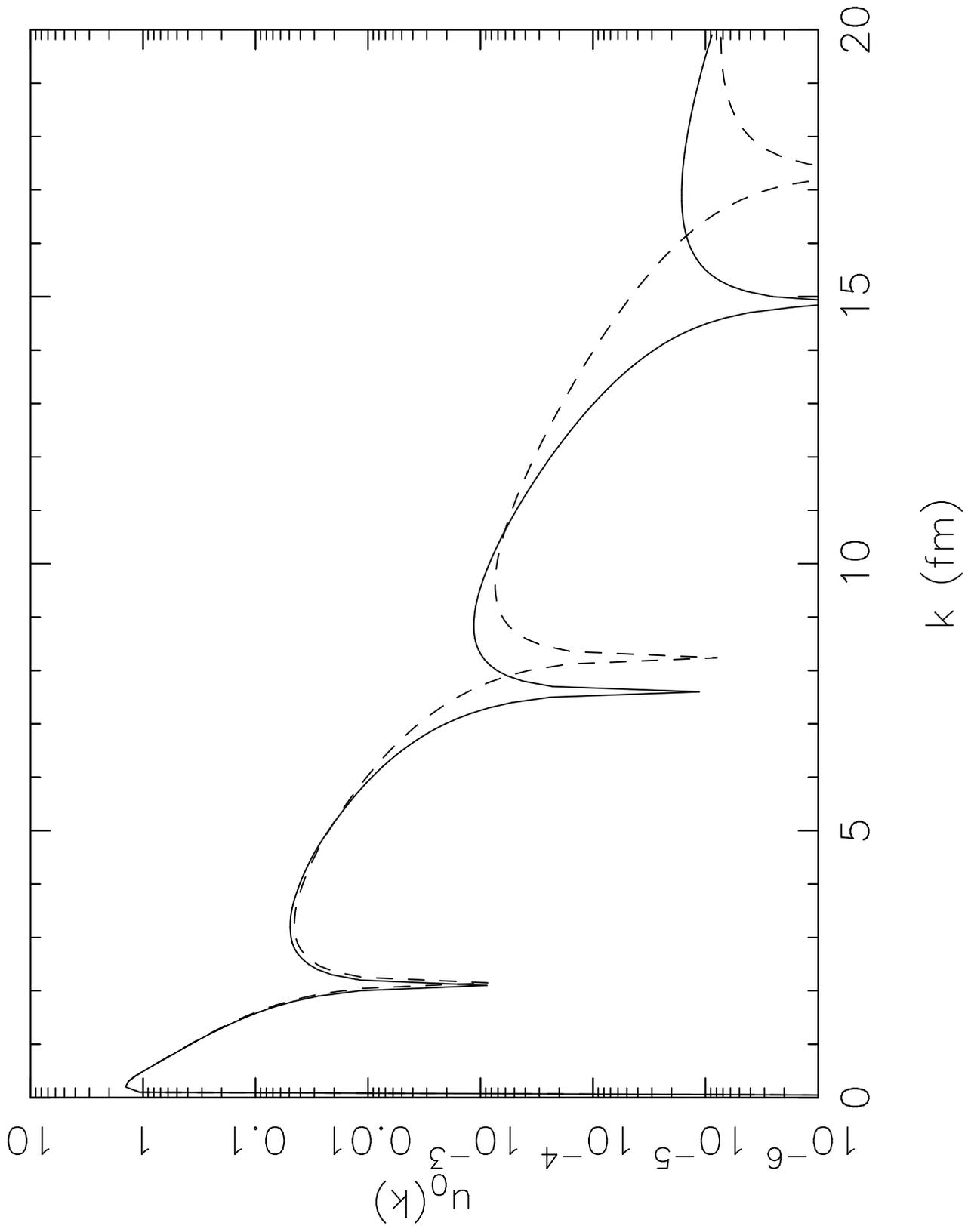
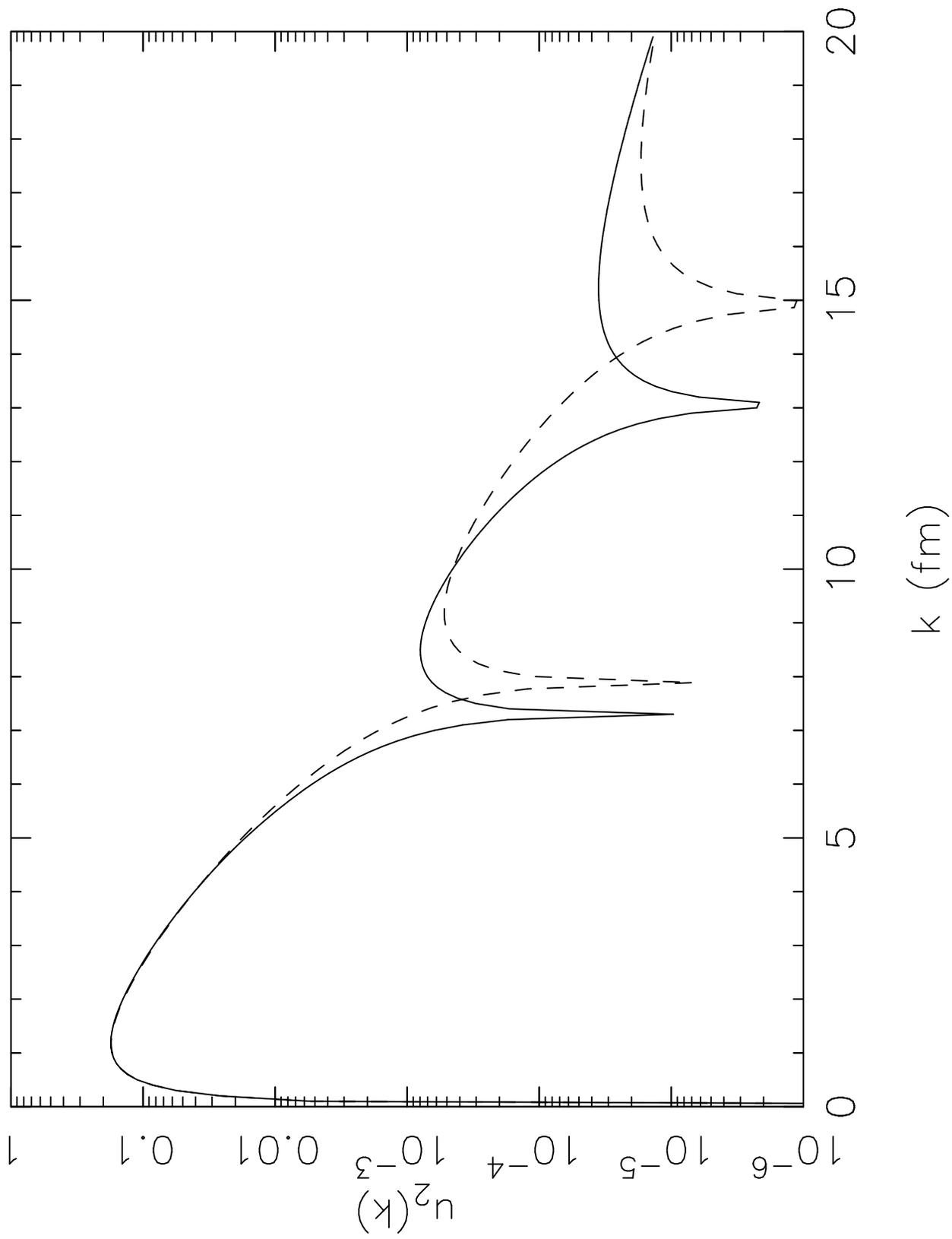
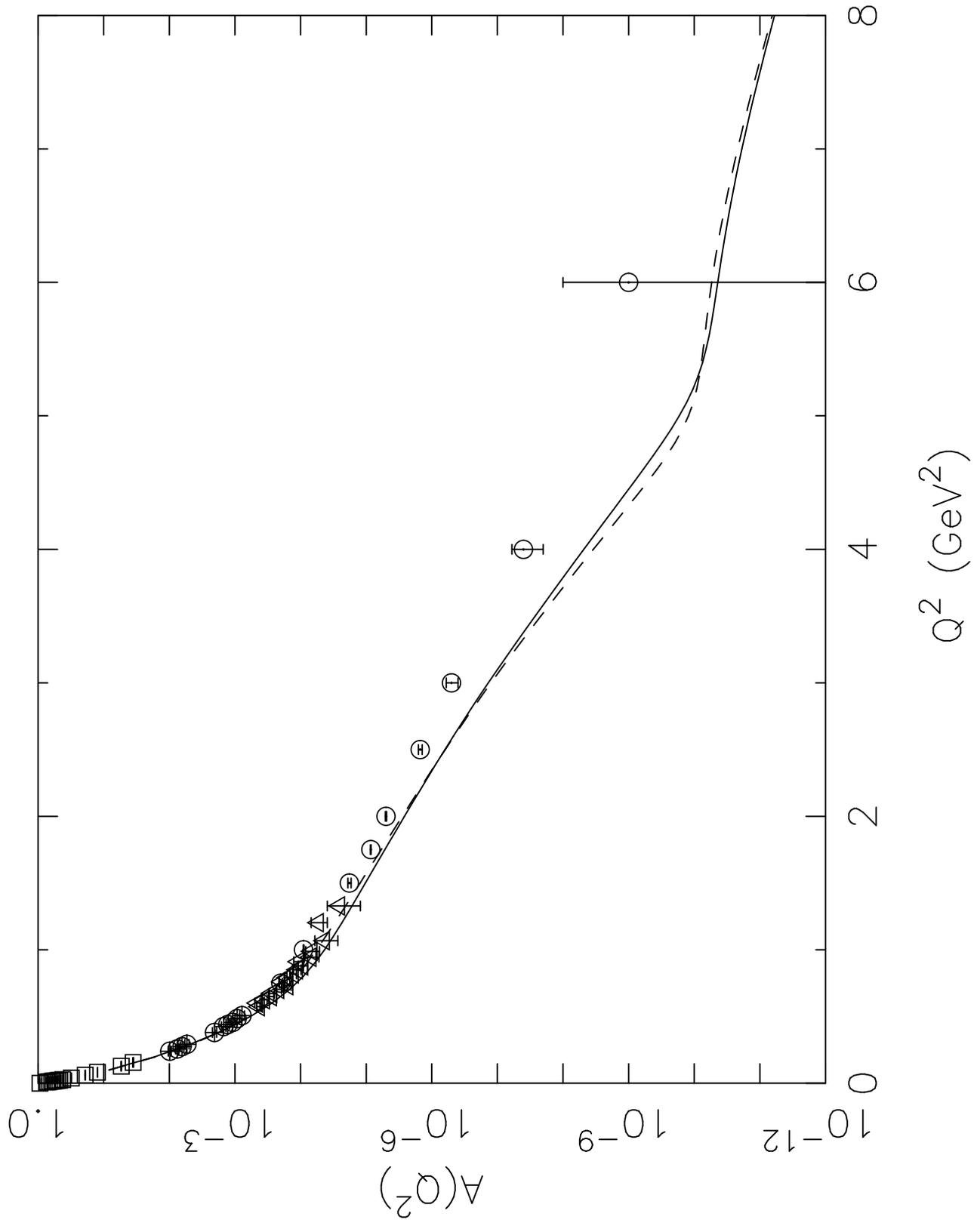
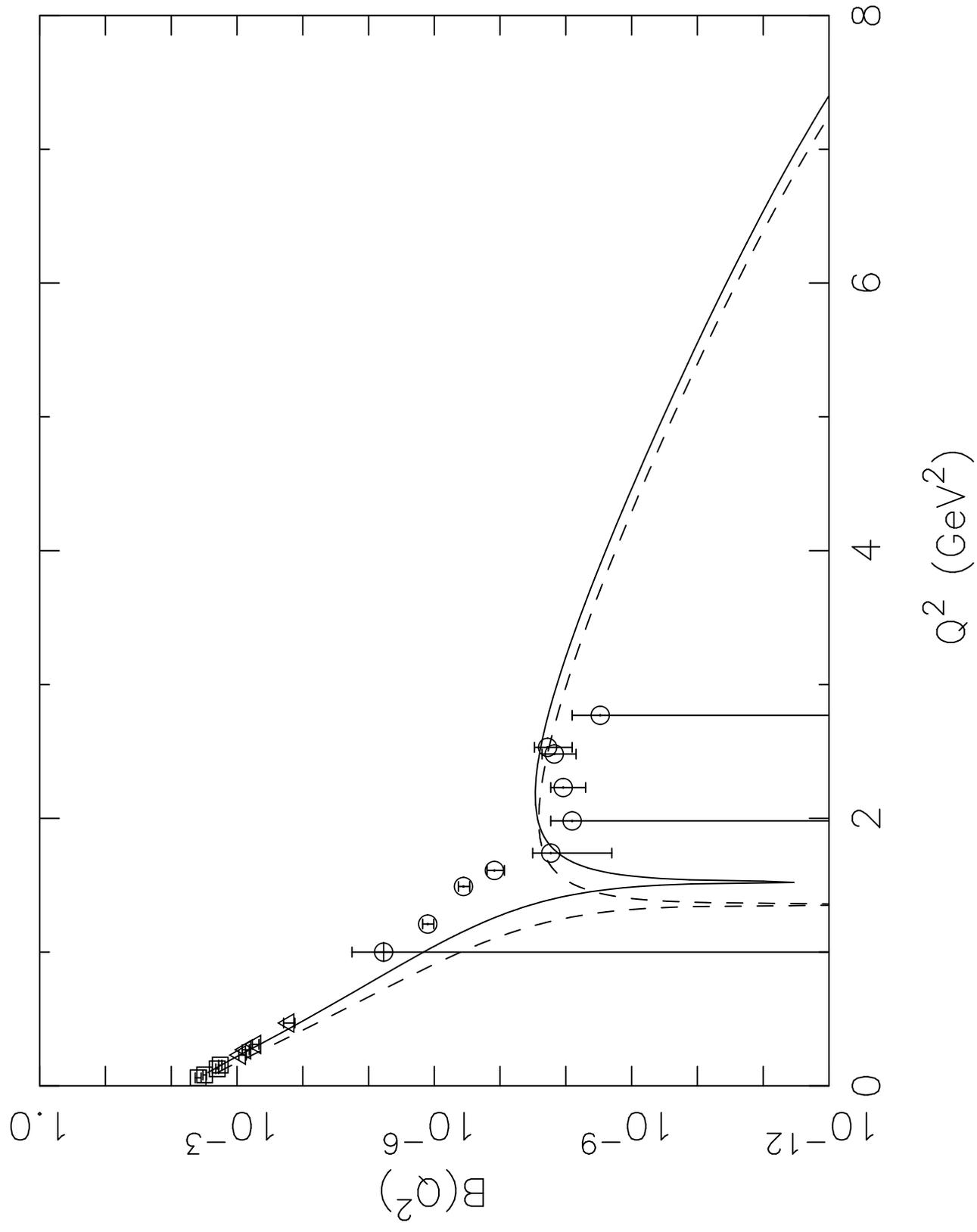
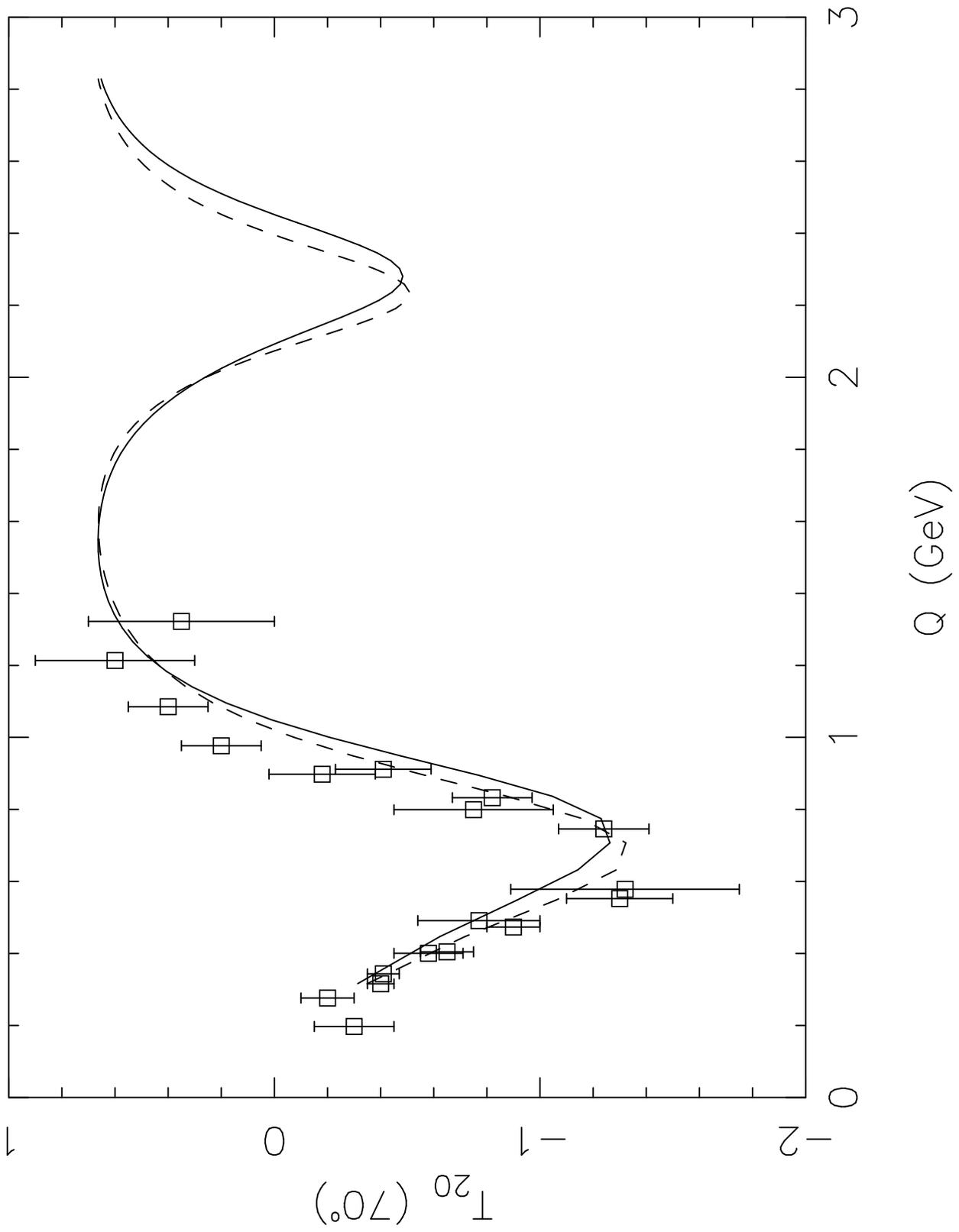
\end{document}